\documentclass[sigconf]{acmart}

\usepackage{booktabs} 

\usepackage{amsmath}
\usepackage{epsfig}
\usepackage{booktabs} 
\usepackage{amssymb}
\usepackage{xspace}

\usepackage{algpseudocode}
\usepackage{algorithm}

\usepackage{color}
\usepackage{subfigure} 
\usepackage{textcomp}

\usepackage{pgfplots}
\usepackage{tikz}

\usepackage{enumitem}
\usepackage{tcolorbox}

\usepackage{xcolor}
\usepackage{listings}

\definecolor{mGreen}{rgb}{0,0.6,0}
\definecolor{mGray}{rgb}{0.5,0.5,0.5}
\definecolor{mPurple}{rgb}{0.58,0,0.82}
\definecolor{backgroundColour}{rgb}{0.95,0.95,0.92}

\lstdefinestyle{CStyle}{
    commentstyle=\color{mGreen},
    keywordstyle=\color{magenta},
    numberstyle=\tiny\color{mGray},
    stringstyle=\color{mPurple},
    basicstyle=\footnotesize,
    breakatwhitespace=false,         
    breaklines=true,                 
    captionpos=b,                    
    keepspaces=true,                 
    numbers=left,                    
    numbersep=5pt,                  
    showspaces=false,                
    showstringspaces=false,
    showtabs=false,                  
    tabsize=2,
    language=C
}

\usepackage{url}

\newcommand{\ie}{i.\nolinebreak[4]\hspace{0.01em}\nolinebreak[4]e.\@\xspace}
\newcommand{\eg}{e.\nolinebreak[4]\hspace{0.01em}\nolinebreak[4]g.\@\xspace}
\newcommand{\etal}{\emph{et al.}\xspace}
\newcommand{\Reals}{{\mathbb R}}

\newcommand{\tsize}{n}

\newcommand{\GPU}{{\sc GPU}}
\newcommand{\CPU}{{\sc CPU}}
\newcommand{\CPUs}{{\sc CPUs}}

\newcommand{\GPUs}{{\sc GPUs}}
\newcommand{\Tset}{\ensuremath{T}}

\newcommand{\minimize}{\operatornamewithlimits{minimize}}

\newcommand{\modelbeta}{\ensuremath{\beta}}

\renewcommand{\vec}[1]{\mathbf{#1}}
\newcommand{\transpose}{^\top}

\newcommand{\norm}[1]{{\|#1\|}}
\usepackage{amsmath}
\usepackage{amssymb}

\newcommand{\ttt}[1]{\texttt{#1}}
\newcommand{\bfast}{BFAST\xspace}
\newcommand{\bfastR}{\ttt{BFAST(R)}\xspace}
\newcommand{\bfastPython}{\ttt{BFAST(Python)}\xspace}
\newcommand{\bfastcpu}{\ttt{BFAST(CPU)}\xspace}
\newcommand{\bfastgpu}{\ttt{BFAST(GPU)}\xspace}

\newcommand{\mosum}{MOSUM\xspace}
\newcommand{\mosums}{MOSUMs\xspace}

\begin{document}
\title{Massively-Parallel Break Detection for Satellite Data}

\author{Malte von Mehren}
\affiliation{%
  \institution{Department of Computer Science, University of Copenhagen}
  \streetaddress{Sigurdsgade 41}
  \city{Copenhagen, Denmark}
}
\email{maltevonmehren@gmail.com}

\author{Fabian Gieseke}
\affiliation{%
  \institution{Department of Computer Science, University of Copenhagen}
  \streetaddress{Sigurdsgade 41}
  \city{Copenhagen, Denmark}
}
\email{fabian.gieseke@di.ku.dk}

\author{Jan Verbesselt}
\affiliation{%
  \institution{Laboratory of Geo-Information Science and Remote Sensing, Wageningen University}
  \streetaddress{Droevendaalse steeg 3}
  \city{Wageningen, The Netherlands}
}
\email{jan.verbesselt@wur.nl}

\author{Sabina Rosca}
\affiliation{%
  \institution{Laboratory of Geo-Information Science and Remote Sensing, Wageningen University}
  \streetaddress{Droevendaalse steeg 3}
  \city{Wageningen, The Netherlands}
}
\email{sabina.rosca@wur.nl}

\author{St\'{e}phanie Horion}
\affiliation{%
  \institution{Department of Geosciences and Natural Resource Management, University of Copenhagen}
  \streetaddress{{\O}ster Voldgade 10}
  \city{Copenhagen, Denmark}
}
\email{stephanie.horion@ign.ku.dk}

\author{Achim Zeileis}
\affiliation{%
  \institution{Department of Statistics,\\University of Innsbruck}
  \streetaddress{Universit\"{a}tsstra\ss{}e 15}
  \city{Innsbruck, Austria}
}
\email{achim.zeileis@uibk.ac.at}

\renewcommand{\shortauthors}{von Mehren et al.}

\begin{abstract} 
The field of remote sensing is nowadays faced with huge amounts of data. While this offers a variety of exciting research opportunities, it also yields significant challenges regarding both computation time and space requirements. In practice, the sheer data volumes render existing approaches too slow for processing and analyzing all the available data. This work aims at accelerating BFAST, one of the state-of-the-art methods for break detection given satellite image time series. In particular, we propose a massively-parallel implementation for BFAST that can effectively make use of modern parallel compute devices such as GPUs. Our experimental evaluation shows that the proposed GPU implementation is up to four orders of magnitude faster than the existing publicly available implementation and up to ten times faster than a corresponding multi-threaded CPU execution. The dramatic decrease in running time renders the analysis of significantly larger datasets possible in seconds or minutes instead of hours or days. We demonstrate the practical benefits of our implementations given both artificial and real datasets.
\end{abstract}

%
%
%


%

\settopmatter{printacmref=false}
\renewcommand\footnotetextcopyrightpermission[1]{}
\pagestyle{plain} 

\maketitle



\section{Introduction}
\label{section:introduction}
The data volumes have increased dramatically in various domains during the last decade. 
A prominent example is the field of remote sensing, which
is nowadays faced with incredible amounts of data that stem from projects such as the \emph{Landsat-8}~\cite{WULDER20122} or the \emph{Sentinel-1} and \emph{Sentinel-2} programs~\cite{rs9090902} producing petabytes of data every year. While this data flood offers the opportunity to address a broad variety of interesting research and industrial applications, it can also make the semi-automatic analysis of all the data extremely time-consuming and, hence, challenging.

An important problem in remote sensing is the task of detecting ``changes'' occurring over time. The key idea is to consider, for the same target region, satellite images at various times and to analyze each of the pixels (or subregions) individually. More precisely, for each pixel, one is essentially given a time series consisting of pixel intensities, which can be used to detect changes over time. One of the state-of-the-art methods for this task is the so-called \emph{break detection for additive season and trend}~(\bfast) approach, which analyzes the pixels individually and generates, for each pixel, an additive season and trend model \textit{``to account for seasonal and trend changes typically occurring within climate-driven biophysical indicators derived from satellite data''}~\cite{Verbesselt201298}. \bfast has been successfully applied for various use cases including, \eg, deforestation~\cite{DeVries2015320,HAMUNYELA2016126} and tropical forest monitoring~\cite{Reiche2016}. However, given the large-scale learning scenarios with images containing millions of individual pixels/regions, such analyses can become very time-consuming, which usually limits the amount of satellite images that can be analyzed.


From a data mining perspective, the \bfast~approach resorts to fitting several regression models for \emph{each} individual pixel---resulting in millions of individual regression models to be fitted for a single image. For analyses covering larger regions of our Earth, \emph{billions} of such regression models have to be fitted per single experiment. This usually results in an extremely time-consuming process that can easily take days or even weeks. Further, this computational bottleneck will become even more significant with future projects that will produce much more data both w.r.t. spatial as well as temporal resolution. Naturally, one way to reduce the practical runtime for this task is to resort to distributed computing frameworks such as Apache Hadoop, see the recent work of Assis~\etal~\cite{AssisQFVLSMC16} for a corresponding implementation. However, depending on the scenario at hand, this might require a large amount of compute resources.

A recent trend in data analytics is to resort to massively-parallel compute devices such as graphics processing units (GPUs) to accelerate such time-consuming tasks~\cite{CatanzaroSK2008,CoatesHWWCN2013,GiesekeHOI2014,GiesekeOI2017,WenZRQT2014}. While such devices offer significant computational resources, the adaptation of existing approaches to the specific needs of these devices can be difficult. In this work, we propose such an adaptation for \bfast. Our experimental evaluation resorts to both artificial and real-world datasets and shows that our massively-parallel scheme is up to four orders of magnitude faster than the commonly used \texttt{R} implementation, up to three orders of  magnitude faster than a direct CPU implementation, and about ten times faster than a corresponding tuned multi-threaded CPU execution. Hence, our implementation can be used to dramatically reduce the practical runtime needed for applying \bfast---rendering it possible to conduct large-scale analyses with hundreds of millions of pixels/time series in minutes or even seconds instead of days using ``cheap'' commodity desktop computers.


This work is structured as follows: In Section~\ref{section:background}, we will outline the details related to the \bfast approach and will also briefly sketch the key principles of massively-parallel programming. The algorithmic framework and the details related to our \GPU~implementation will be described in Section~\ref{section:algorithmic_framework}, followed by an experimental evaluation provided in Section~\ref{section:experiments}. Conclusions are drawn in Section~\ref{section:conclusions}.

\section{Background}
\label{section:background}
We start by outlining the key principles of the \bfast approach, followed by a quick introduction to basic concepts of massively-parallel programming.

\subsection{Break Detection}
\label{sec:break_detection}
\begin{figure}[t]
\begin{center}
\resizebox{0.98\columnwidth}{!}{\includegraphics{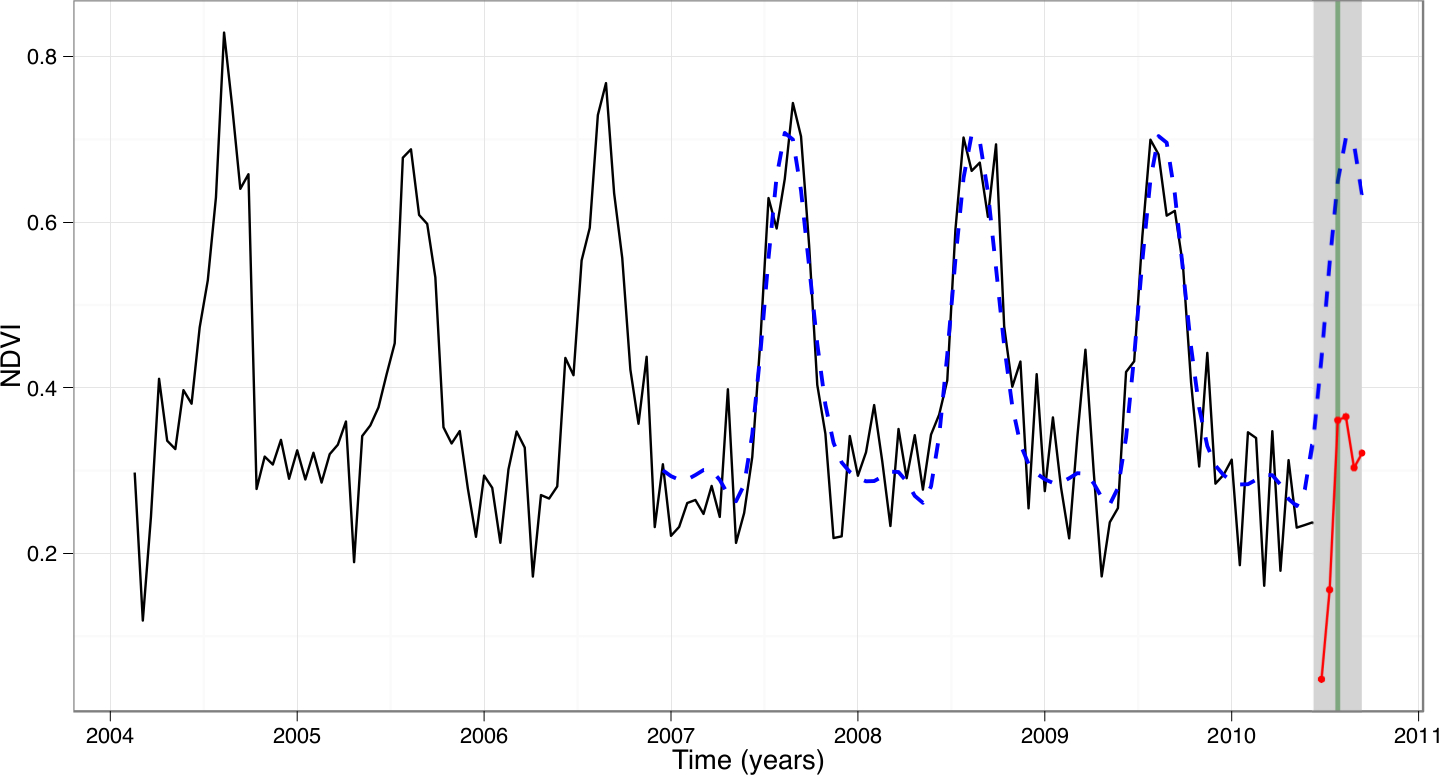}} 
\end{center}
\caption{Automatic break detection via \bfast~\cite{Verbesselt201298}: The dotted blue line corresponds to the predicted values, but the time series (red line) does not follow the model anymore. Here, \bfast detects a break in the time series (gray box). The time series data are based on the so-called NDVI index, which captures the amount of green vegetation for a given target pixel/region.}
\label{fig:bfast}
\end{figure}

The task of break detection has various applications in remote sensing. In this work, we focus on the \emph{\bfast (monitor)} approach introduced by Verbesselt~\etal~\cite{Verbesselt201298}, which depicts one of the most prominent techniques for this problem. Typically, one makes use of so-called spectral vegetation indices such as the \emph{Normalised Difference Vegetation Index}~(NDVI)~\cite{MyneniHSM95} to extract, for each pixel, a single value from multi-spectral satellite image data (reflecting the amount of ``green'' vegetation at that pixel). Given multiple images for the same region taken at different times, this gives rise to a time series $y_1, \ldots, y_N$ for each individual pixel, see Figure~\ref{fig:bfast} for an illustration. 

The basic idea of \bfast is to assume a season-trend model with linear trend and harmonic season. More specifically, the time series data are modeled via
\begin{equation}
\label{eq:modelpure}
y_t = \alpha_1 + \alpha_2t + \sum\limits_{j=1}^k \gamma_j \sin \left( \frac{2 \pi j t}{f} + \delta_j \right) + \epsilon_t,
\end{equation}
where the unknown parameters are the \emph{intercept} $\alpha_1$, the \emph{slope}~$\alpha_2$ (trend), the \emph{amplitudes} $\gamma_1, ... ,\gamma_k$, and the \emph{phases} $\delta_1, ... , \delta_k$ (\ie, seasons). Furthermore, $f$ is the \emph{frequency} of the observations (\eg, $f=23$ observations per year for a time series with an interval of 16 days between the different observations) and $k$ is the number of \emph{harmonic terms} that capture the seasonal pattern (\eg, $k = 3$). Finally, $\epsilon_t$ depicts the unobservable error at time $t=1, \ldots, N$. 

The model~(\ref{eq:modelpure}) can be written as a standard linear model having the form 
\begin{equation}
\label{eq:modelregression}
 y_t = \vec{x}_t\transpose \modelbeta + \epsilon_t
\end{equation}
with 
\begin{equation*}
\vec{x}_t = \left(1, t, \sin(F(1)), \cos(F(1)), \ldots, \sin(F(k)), \cos(F(k)) \right)\transpose
\end{equation*}
and
\begin{equation*}
\resizebox{0.99\columnwidth}{!}{$
\modelbeta = \left( \alpha_1 , \alpha_2 , \gamma_1 \cos \left( \delta_1 \right) , \gamma_1 \sin \left( \delta_1 \right) , ... , \gamma_k \cos \left( \delta_k \right) , \gamma_k \sin \left( \delta_k \right) \right)\transpose$}
\end{equation*}
where $F(j) = \frac{2 \pi j t}{f}$, see Cryer and Chan~\cite[Section 3.3]{CryerC2008}. 


The basic idea of \bfast is to split the available time series data into two parts: The first one, called \emph{stable history period}, consists of the first $n$ elements of the time series, \ie, $y_1, \ldots, y_n$. It is assumed to be known in advance and is utilized to consistently estimate the parameter vector $\hat \modelbeta$ and the error variance $\hat \sigma^2$ via least squares. The second part, $y_{n+1}, \ldots, y_N$, is called the \emph{monitor period} that is tested for ``breaks'' (i.e., changes in the regression coefficients over time) by comparing the observed data with the predictions from the stable history model, see again Figure~\ref{fig:bfast} for an illustration.

To measure the discrepancy between the model and the measurements for the monitor period, \bfast resorts to a \emph{moving sums} (\mosum) process with \emph{bandwidth} $1 \leq h \leq n$ defined for the monitor period $t=n+1,\ldots,N$:
\begin{equation}
\label{eq:mosum}
MO_t = \frac{1}{\hat{\sigma}\sqrt{n}} \sum\limits_{s=t-h+1}^t \left( y_s - \vec{x}_s\transpose \hat{\modelbeta} \right)
\end{equation}
Since we assumed a stable history period, the model should stay stable in the monitor period if no breaks occur. This means that under this assumption of structural stability, the \mosum process will only fluctuate randomly around zero. In case of a break, however, it should systematically deviate away from zero, and a break will be declared if the \mosum process \textit{``exceeds some boundary that is asymptotically only crossed with 5\% probability''}~\cite{Verbesselt201298}. For a time $t$, the boundary $b_t$ is defined via
\begin{equation}
\label{eq:boundary}
b_t  = \lambda \sqrt{\log_{+}{\frac{t}{n}}}
\end{equation}
with 
\begin{equation*}
 \log_{+}{x} = \begin{cases}
                 1       & x \leq e\\
                 \log{x} & otherwise
                \end{cases}.
\end{equation*}
Here, $\lambda$ is the critical value chosen such that a random boundary crossing occurs with probability $\alpha$. In addition to $\alpha$, the value of~$\lambda$ also depends on~$h$ and the \emph{monitoring horizon}~$N/n$. The specific value of $\lambda$ has been found by simulation of different values of $\alpha$, $h$, and $N/n$~\cite{Verbesselt201298}.


From a computational perspective, one generates, for each time series, a training set $\Tset = \{(\vec{x}_1,y_1),\ldots,(\vec{x}_N,y_N)\} \subset \Reals^{2 +2k} \times \Reals$ that is used to fit and apply a standard ordinary least-squares model. More precisely, in the first step, one solves
\begin{align}
\label{eq:ols}
\minimize_{\modelbeta \in \Reals^{2+2k}} &\; \norm{\vec{y}_{[:n]} - {\vec{X}_{[:,:n]}}\transpose {\modelbeta}}_2^2
\end{align}
with $\vec{y}_{[:n]} \in \Reals^{n \times 1}$ containing the first $n$ observations and $\vec{X}_{[:,:n]} \in \Reals^{{(2+2k)} \times \tsize}$ containing the first $n$ columns of $\vec{X}$ consisting of the vectors $\vec{x}_1, \ldots, \vec{x}_N$ as columns.\footnote{In the following, we make use of the slicing operator with $\vec{y}_{[:n]}$ taking the first $n$ elements of the vector $\vec{y}$, $\vec{X}_{[:,:n]}$ taking the first $n$ columns of the matrix $\vec{X}$, and $\vec{X}_{[:n,:]}$ taking the first $n$ rows of the matrix $\vec{X}$.} A solution to this task can be obtained via 
\begin{equation}
\label{eq:ols-model}
 \hat{\modelbeta} = (\vec{X}_{[:,:n]} {\vec{X}_{[:,:n]}}\transpose)^{-1} \vec{X}_{[:,:n]} \vec{y}_{[:n]}
\end{equation}
where $(\vec{X_{[:,:n]}} {\vec{X}_{[:,:n]}}\transpose)^{-1} \vec{X}_{[:,:n]}$ is the \emph{Moore-Penrose pseudo-inverse} of~$\vec{X}_{[:,:n]}$~\cite{HastieTF2009}.
Afterwards, in the second step, one makes use of this model to compute the \mosum process and the breakpoints for the monitoring period via Equations~(\ref{eq:mosum}) and~(\ref{eq:boundary}). 

\begin{algorithm}[t]
  \caption{\bfast}
  \label{alg:bfast}
  \begin{algorithmic}[1]
    \Require
      \Statex Time series as vector $\vec{y} = (y_1, \ldots, y_N)\transpose$
      \Statex Annual frequency $f$
      \Statex Size of history period $n, 1 \leq n < N$
      \Statex Width of \mosum window $h, 1 \leq h \leq n$
      \Statex Number of harmonic terms $k$
      \Statex Significance level $\alpha$
    \Ensure 1D array \ttt{D} (bools) containing detected breaks
    \State $\vec{X} = \begin{bmatrix}
                                            1                             & \dots  & 1                               \\
                                            1                             & \dots  & N                               \\
                                            \sin\left(2 \pi 1 / f \right) & \dots  & \sin\left(2 \pi N / f \right)   \\
                                            \cos\left(2 \pi 1 / f \right) & \dots  & \cos\left(2 \pi N / f \right)   \\
                                            \vdots                        & \ddots & \vdots                          \\
                                            \sin\left(2 \pi k / f \right) & \dots  & \sin\left(2 \pi k N / f \right) \\
                                            \cos\left(2 \pi k / f \right) & \dots  & \cos\left(2 \pi k N / f \right) \\
                                          \end{bmatrix}$ \Comment{$O(Nk)$}

    \State $\hat{\modelbeta} = (\vec{X}_{[:,:n]} {\vec{X}_{[:,:n]}}\transpose)^{-1} \vec{X}_{[:,:n]} \vec{y}_{[:n]}$ \Comment{$O(k^3 + k^2n)$}
    \State $\hat{\vec{y}} = \vec{X}\transpose\hat{\modelbeta}$  \Comment{$O(Nk)$}
    \State $\vec{r} = \hat{\vec{y}} - \vec{y}$ \Comment{$O(N)$}
    \State $\hat{\sigma} = \sqrt{\frac{\sum_{i=1}^{n}r_i^2}{n-(2+2k)}}$ \Comment{$O(n)$}
    \For{$t = n+1, \dots, N$} 
      \State $\ttt{MO[t-n-1]} = \frac{1}{\hat{\sigma} \sqrt{n}} \cdot \sum_{i=t-h}^{t} r_i$ \Comment{$O(h)$}
    \EndFor
    \State Let $\lambda$ be critical value (based on $h$, $N/n$, and $\alpha$) \Comment{$O(1)$}
    \For{$t = n+1, \dots, N$} 
      \State $\ttt{BOUND[t-n-1]} = \lambda \sqrt{\log_{+}{\frac{t}{n}}}$ \Comment{$O(1)$}
    \EndFor
    \State $\ttt{D} = \lvert \ttt{MO} \rvert > \ttt{BOUND}$ \Comment{$O(N-n)$}
  \end{algorithmic}
\end{algorithm}
The overall approach along with associated runtimes are given in Algorithm~\ref{alg:bfast}: The matrix $\vec{X} \in \Reals^{{(2+2k)} \times N}$ is initialized in Step~1. Step~2 computes the linear model based on the history period and Step 3 generates the predictions for the entire time series. 
The \mosum process is calculated in Steps 4--8, and the final steps take care of detecting potential breaks.

Note that fitting the model and computing the \mosum process/breaks are computationally not very demanding given a \emph{single} time series. However, for a typical scenario, the computations outlined in Algorithm~\ref{alg:bfast} have to be conducted for millions of time series---leading to a very time-consuming task even given moderate-sized datasets. More precisely, given $m$ time series, applying \bfast takes $\mathcal{O}(m \cdot (k^3 + k^2n + Nk))$ runtime in total.

\subsection{Massively-Parallel Computing}
We will focus on \emph{graphics processing units}~(\GPUs) as many-core devices in this work. In their original form, such devices have been exclusively used for accelerating computer graphics, but today's \GPUs~are also well suited for general computations such as matrix-matrix multiplications, which led to the concept of \emph{general-purpose computing on graphics processing units}~(GPGPU). In contrast to multi-core processors, which usually resort to a small number of ``complex'' compute units, graphics processing units often contain thousands of ``simple'' compute units: A standard \CPU~execution is based on complex control units and mechanisms that are tuned for a sequential execution of programs. In contrast, \GPUs~are based on much simpler control units and are optimized for code being executed in a massively-parallel fashion~\cite{ChengGM2014}. More precisely, threads are executed in parallel based on the \emph{single instruction multiple data}-paradigm, which means that all threads belonging to a thread group (typically a set of 32 threads) have to execute the same instruction in a single clock cycle, but can access different locations in memory. This depicts a restriction compared to a more complex \CPU~execution and not all programs might be executable this way. However, in case it is possible to execute a program in such a manner, \GPUs~usually achieve a much higher performance compared to \CPUs.

Graphics processing units and other many-core devices offer huge computational resources and have been successfully applied in data analysis in the past years, see, \eg,~\cite{CatanzaroSK2008,CoatesHWWCN2013,GiesekeHOI2014,GiesekeOI2017,WenZRQT2014}. A general goal of such implementations is to conduct the ``inexpensive'' computations via the \CPU~and the ``expensive'' ones using the \GPU. To achieve efficiency, the algorithmic workflows of the algorithms at hand usually have to be adapted such that they are suited for a massively-parallel code execution. Two general principles are important in this context: The first one is that sufficient parallelism has to be available to make fully use of a \GPU, \ie, it must be possible to split the compute intensive parts into thousands of individual subtasks. The second main principle is an efficient access to the memory of the device as well as of the host (e.g., minimizing memory transfers between host and device).

\section{Algorithmic Framework}
\label{section:algorithmic_framework}

While being conceptually very easy to implement, a direct implementation of \bfast is not very efficient and usually leads to significant practical runtimes even for moderate-sized datasets. In this section, we derive both an efficient CPU and an GPU implementation. We start by outlining the efficient multi-core implementation of the \bfast approach, followed by the modifications needed to obtain an efficient massively-parallel version.

The \bfast approach as described in Algorithm~\ref{alg:bfast} considers the time series individually, \ie, model fitting and break detection are conducted individually per time series corresponding to a single pixel of the image scene at hand. In the following, we will focus on learning scenarios with (almost) complete time series data for all pixels of a given scene.\footnote{In case of almost complete time series, one can, e.g., resort to simple schemes such as forward/backward filling to remove the missing values (spending linear time).} In this case, all time series can be written as a matrix $\vec{Y} \in \Reals^{N \times m}$ having the form 
\begin{equation}
\begin{split}
   \vec{Y} & = \begin{bmatrix}
          y_{1,1}   & \dots  & y_{1,m} \\
          \vdots    & \ddots & \vdots \\
          y_{N,1}   & \dots  & y_{N,m} \\
         \end{bmatrix}
         .
\end{split} 
\end{equation}
Given the matrix $\vec{Y}$, one can combine many of the computations conducted for the individual pixels. More precisely, one can fuse the operations conducted in Steps~1 and 2 of Algorithm~\ref{alg:bfast} by computing 
\begin{equation}
\vec{M} = (\vec{X}_{[:,:n]} \vec{X}_{[:,:n]}\transpose)^{-1} \vec{X}_{[:,:n]} 
\end{equation}
only once for all time series. Given the matrix $\vec{M} \in \Reals^{{(2+2k)} \times n}$, one can obtain optimal coefficients for (\ref{eq:ols-model}) for all time series via
\begin{equation}
\hat{\modelbeta}_{all} = \left(\hat{\modelbeta}_1, ..., \hat{\modelbeta}_m\right)\transpose = \vec{M} \vec{Y}_{[:n,:]} \in \Reals^{2+2k \times m}.
\end{equation}
All predictions and the associated residuals are then given by 
\begin{equation}
 \hat{\vec{Y}} = \left(\hat{y}_1, ..., \hat{y}_m\right)\transpose = \vec{X}\transpose \hat{\modelbeta}_{all}
\end{equation}
and 
\begin{equation}
 \vec{R} = \vec{Y} - \hat{\vec{Y}},
\end{equation}
respectively. All operations described above are based on matrix operations, which are well-suited for a multi-threaded execution.
The remaining computations are related to the \mosum processes and the detection of breaks. Here, efficient multi-threaded implementations can be obtained by parallelizing over the $m$ time series using, \eg, \texttt{OpenMP}~\cite{Dagum:1998:OIA:615255.615542}. Note that one can also update the computations of the partial sums in an efficient parallel manner (Step~7 of Algorithm~\ref{alg:bfast}).
%
%
%
%
In total, one needs $\mathcal{O}(k^3 + k^2 (n+m) + Nk)$ time for processing (images containing) $m$ time series. This is about $\mathcal{O}(N)$ times faster than a direct implementation of Algorithm~\ref{alg:bfast}. Also, the remaining compute intensive parts (involving the $m$ individual time series) can be efficiently parallelized. Hence, to sum up, one can obtain an efficient multi-core implementation by the steps outlined above.
\begin{algorithm}[t]
  \caption{\bfastgpu}
  \label{alg:bfast-gpu}
  \begin{algorithmic}[1]
    \Require
      \Statex Matrix $\vec{Y}$ containing all time series
      \begin{equation*}
	    \ttt{Y} = \begin{bmatrix}
		  y_{1,1}   & \dots  & y_{1,m} \\
		  \vdots    & \ddots & \vdots \\
		  y_{N,1}   & \dots  & y_{N,m} \\
		  \end{bmatrix}
      \end{equation*}
      \Statex Annual frequency $f$
      \Statex Size of history period $n$, $1 \leq n < N$
      \Statex Width of \mosum window $h$, $1 \leq h \leq n$
      \Statex Number of harmonic terms $k$
      \Statex Significance level $\alpha$
    \Ensure 1D array \ttt{D} (bools) containing detected breaks
    \State 
    $\vec{X} = \begin{bmatrix}
                                            1                             & \dots  & 1                               \\
                                            1                             & \dots  & N                               \\
                                            \sin\left(2 \pi 1 / f \right) & \dots  & \sin\left(2 \pi N / f \right)   \\
                                            \cos\left(2 \pi 1 / f \right) & \dots  & \cos\left(2 \pi N / f \right)   \\
                                            \vdots                        & \ddots & \vdots                          \\
                                            \sin\left(2 \pi k / f \right) & \dots  & \sin\left(2 \pi k N / f \right) \\
                                            \cos\left(2 \pi k / f \right) & \dots  & \cos\left(2 \pi k N / f \right) \\
                                          \end{bmatrix}$ 
    \State Transfer $\vec{X}$ and $\vec{Y}$ to device \Comment{$O(Nk + Nm)$ transfer}
    \State Compute $\vec{M} = {(\vec{X}_{[:,:n]} {\vec{X}_{[:,:n]}}\transpose)}^{-1} \vec{X}_{[:,:n]}$ 
    \State Compute $\hat{\modelbeta}_{all} = \vec{M} \vec{Y}[:n,:]$ 
    \State $\hat{\vec{Y}} = \vec{X}\transpose \hat{\modelbeta}_{all}$
    \State Allocate device memory for $(N-n) \times m$ array \ttt{MO} 
    \State $\ttt{moving\_sums} \left(\ttt{MO}, \ttt{Y}, \ttt{\^{Y}}, n-h, h, n, N-n, m \right)$ 
    \State Allocate host memory for array \ttt{BOUND} of size $N-n$
    \For{$t = n+1, \dots, N$} 
      \State $\ttt{BOUND[t-n-1]} = \lambda \sqrt{\log_{+}{\frac{t}{n}}}$ 
    \EndFor
    \State Transfer \ttt{BOUND} to device \Comment{$O(N-n)$ transfer}
    \State Allocate device memory for array \ttt{D} of size $m$
    \State $\ttt{detect\_breaks} \left(\ttt{MO}, \ttt{BOUND}, \ttt{D}, h, n \right)$ 
    \State Transfer \ttt{D} to host \Comment{$O(m)$ transfer}
  \end{algorithmic}
\end{algorithm}
The modifications needed to obtain an efficient many-core variant are shown in Algorithm~\ref{alg:bfast-gpu}: In Steps 1 and 2, $\vec{X}$ and $\vec{Y}$ are instantiated and transferred to device memory. Here, the $\vec{Y}$ clearly dominates the amount of data that is transferred from host to device in case of large $m$ (many time series). After transferring the data, all models are computed in Steps 3 and 4 on the device, followed by the computation of the predictions in Step 5. For both phases, the operations involving $\vec{Y}$ usually dominate the runtime---and these phases greatly benefit from efficient many-core matrix libraries provided by, \eg, \texttt{CUDA}. 

The following steps compute the residuals, moving sums, and breaks in a massively-parallel fashion. The overall efficiency also depends on an efficient implementation of these steps. The \ttt{CUDA} kernel used for \mosum is shown in Algorithm~\ref{alg:moving_sums_gpu}. In total, $m$ threads are spawned, one thread for each \mosum process. Each thread first computes an initial sum (Lines 17--21), which is then updated to obtain all sums (Lines 22--27). The final \mosum values are computed in Lines 28--39. Note that the residuals are recomputed on the fly for all steps to save device memory. This trade-off was chosen since the computational parts of the implementation only constitutes a small part of the overall runtime, while the phase of transferring the data from host to device memory takes up the vast majority (see Section~\ref{section:experiments}). Note that all threads within a warp execute the same operations. In addition, the involved arrays are accessed in a transposed manner, leading to coalesced memory access pattern.


Finally, the breaks detected in Step 14 are transferred back to host (the transfer time for this step is significantly smaller than the one for moving $\vec{Y}$ from host to device). We only have to transfer the breaks back from the GPU, not the intermediary results, even though these are available on the CPU version. If one wants to analyze the residuals, \mosum, or another intermediary result in a specific time series, one can perform the analysis on the CPU for these specific time series after learning where the breaks are from the \GPU~analysis.


\begin{algorithm}[t]
  \caption{\ttt{moving\_sums}}
  \label{alg:moving_sums_gpu}
\begin{lstlisting}[style=CStyle]
moving_sums(float *MO, 
	    float *Y, 
	    float *YH, 
	    int s, 
	    int ws, 
	    int hs, 
	    int ms, 
	    int m, 
	    int k)
	    {

  uint tid = threadIdx.x;
  uint gid = blockIdx.x*blockDim.x+tid;

  if (gid < m) {
    int j;
    // compute initial sum
    MO[gid]=0.0;
    for (j=s+1; j<s+ws+1; j++) {
      MO[gid] += (Y[gid+j*m]-YH[gid+j*m]);
    }
    // compute remaining sums (updates)
    for (j=1; j<ms; j++) {
      MO[gid+j*m] = MO[gid+(j-1)*m]-
       (Y[gid+(s+j)*m]-YH[gid+(s+j)*m])+
       (Y[gid+(s+ws+j)*m]-YH[gid+(s+ws+j)*m]);
    }
    // compute variance
    float sigma=0.0;
    for (j=0; j<hs; j++) {
      sigma += (Y[gid+j*m]-YH[gid+j*m])* 
               (Y[gid+j*m]-YH[gid+j*m]);
    }
    sigma = sqrt(sigma/(hs-(2+2*k)));
    // final mosum values
    for (j=0; j<ms; j++) {
      float tmp = (sigma*sqrt((float)hs));
      MO[gid+j*m] = MO[gid+j*m]/tmp;
    }
  }
}
\end{lstlisting}
\end{algorithm}





%
%
%
%

\section{Experiments}
\label{section:experiments}
We conduct various runtime experiments using artificial datasets to analyze the practical runtimes and the achieved speed-ups. We also consider a large-scale satellite image time series dataset to sketch the benefits of our new many-core implementation.

\subsection{Setup}
For all runtime experiments, we resort to a standard commodity desktop computer with an \ttt{Intel(R) Core(TM) i5-4460} CPU at 3.20GHz (4 cores, 4 hardware threads), 8 GB RAM, and a \ttt{GeForce GTX 790} \GPU~(4 GB RAM) with \ttt{Ubuntu 16.04} (64 bit) as operating system.\footnote{These hardware specifications were chosen since this analysis is most commonly performed on regular desktop computers, making the analysis a more realistic benchmark for real-world use of \bfast.} All implementations resort to \ttt{Python 2.7} along with \ttt{C} code compiled using \ttt{Swig} with \ttt{gcc-5.4.0} and \ttt{-fopenmp} as additional compiler options. All matrix operations are conducted via the \texttt{Numpy} package (compiled against efficient matrix libraries). For the \GPU~implementation, we make use of the \emph{scikit-cuda} package~\cite{givon_scikit-cuda_2015} and of custom \texttt{CUDA} kernels, called from \texttt{Python} via the \emph{PyCuda} package~\cite{kloeckner_pycuda_2012}.

We will consider the following four implementations for all runtime experiments:
\begin{enumerate}
  \item \bfastR: The available \ttt{R} implementation of \bfast that is commonly used in remote sensing for conducting \bfast~analyses.
  \item \bfastPython: A direct implementation of~\bfast as shown in Algorithm~\ref{alg:bfast} using \ttt{Python}, where the \emph{Numpy} package is used for all compute-intensive parts and the \emph{Scikit-Learn}~\cite{scikit-learn} package for computing the linear models. 
  \item \bfastcpu: The multi-core implementation of \bfast as described in Section~\ref{section:algorithmic_framework}. All compute-intensive parts are either conducted via the \emph{Numpy} package or via \emph{Swig} along with corresponding \ttt{C} implementations. This implementation can be seen as the direct competitor of the many-core implementation. Note that all involved parts benefit from the multi-core execution: The matrix-based operations are executed via the routines provived by the \emph{Numpy} package (linked against Atlas/BLAS). The remaining parts are parallelized over the number $m$ of pixels via \texttt{OpenMP}.
  \item \bfastgpu: The many-core implementation proposed in this work as described in Section~\ref{section:algorithmic_framework}. Here, all compute-intensive parts are either conducted via the \ttt{CUDA} matrix libraries (using \emph{scikit-cuda}) or via direct calls of \ttt{CUDA} kernels.
\end{enumerate}

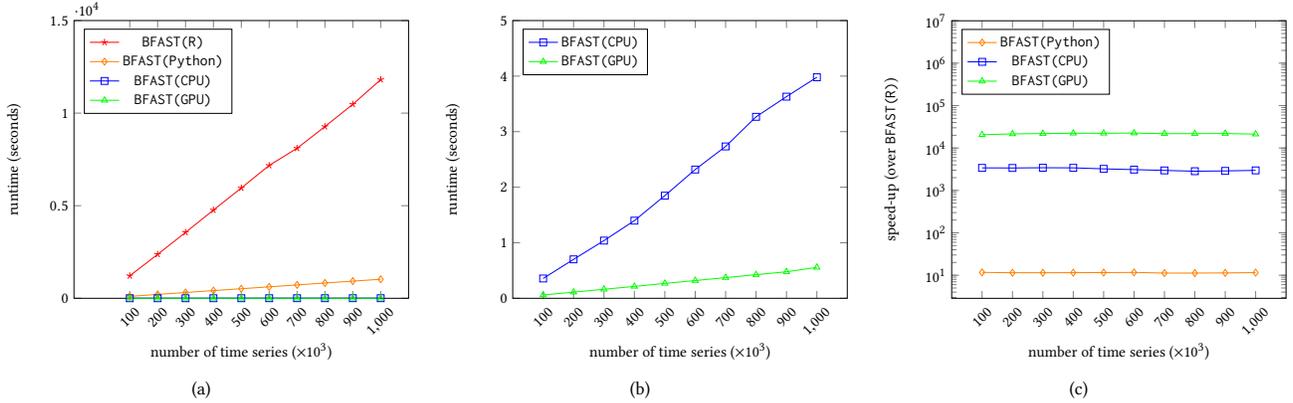
\begin{figure*}[t]
 \centering
 \begin{subfigure}[]{
  \centering
  \resizebox{0.31\textwidth}{!}{
  \begin{tikzpicture}
   \begin{axis}[
       xlabel={number of time series ($\times 10^3$)},
       ylabel={runtime (seconds)},
       xmin=-100, xmax=1100,
       ymin=-1, ymax=15000,
       xtick={100,200,300,400,500,600,700,800,900,1000},
       legend pos=north west,
       xticklabel style={rotate=45},
       x label style={at={(axis description cs:0.5,-0.05)},anchor=north},
   ]
    
   \addplot[color=red,mark=star,]
       coordinates {(100,1211.8134)
                    (200,2371.9806)
                    (300,3557.3364)
                    (400,4761.9864)
                    (500,5954.4972)
                    (600,7173.2952)
                    (700,8100.5292)
                    (800,9272.2608)
                    (900,10476.9576)
                    (1000,11807.4636)};
    
   \addplot[color=orange,mark=diamond,]
       coordinates {(100,103.675951958)
                    (200,207.420264006 )
                    (300,310.846486092)
                    (400,414.381913185)
                    (500,515.39418602)
                    (600,615.628947973)
                    (700,720.526872873)
                    (800,824.292226076)
                    (900,925.139184952)
                    (1000,1025.9081111)};
    
   \addplot[color=blue,mark=square,]
       coordinates {(100,0.35619020462)
                    (200,0.700788021088)
                    (300,1.03847789764)
                    (400,1.3975892067)
                    (500,1.84744691849)
                    (600,2.31781291962)
                    (700,2.7338809967)
                    (800,3.26691794395)
                    (900,3.63009190559)
                    (1000,3.97938990593)};
    
   \addplot[color=green,mark=triangle,]
       coordinates {(100,0.0588018894196)
                    (200,0.11029791832)
                    (300,0.162318944931)
                    (400,0.213392019272)
                    (500,0.267430067062)
                    (600,0.318933010101)
                    (700,0.371750831604)
                    (800,0.424654006958)
                    (900,0.477159976959)
                    (1000,0.5558822155)};
   \legend{\bfastR,\bfastPython,\bfastcpu,\bfastgpu}
   \end{axis}
  \end{tikzpicture}
  }
  }
 \end{subfigure}
 \begin{subfigure}[]{
  \centering
  \resizebox{0.31\textwidth}{!}{
  \begin{tikzpicture}
   \begin{axis}[
       xlabel={number of time series ($\times 10^3$)},
       ylabel={runtime (seconds)},
       xmin=0, xmax=1100,
       ymin=0, ymax=5,
       xtick={100,200,300,400,500,600,700,800,900,1000},
       legend pos=north west,
       xticklabel style={rotate=45},
       x label style={at={(axis description cs:0.5,-0.05)},anchor=north},
   ]
    
   \addplot[color=blue,mark=square,]
       coordinates {(100,0.35619020462)
                    (200,0.700788021088)
                    (300,1.03847789764)
                    (400,1.3975892067)
                    (500,1.84744691849)
                    (600,2.31781291962)
                    (700,2.7338809967)
                    (800,3.26691794395)
                    (900,3.63009190559)
                    (1000,3.97938990593)};
    
   \addplot[color=green,mark=triangle,]
       coordinates {(100,0.0588018894196)
                    (200,0.11029791832)
                    (300,0.162318944931)
                    (400,0.213392019272)
                    (500,0.267430067062)
                    (600,0.318933010101)
                    (700,0.371750831604)
                    (800,0.424654006958)
                    (900,0.477159976959)
                    (1000,0.5558822155)};

   \legend{\bfastcpu,\bfastgpu}
   \end{axis}
  \end{tikzpicture}
  }
  }
 \end{subfigure}
 \begin{subfigure}[]{
  \centering
  \resizebox{0.31\textwidth}{!}{
  \begin{tikzpicture}
   \begin{axis}[
       xlabel={number of time series ($\times 10^3$)},
       ylabel={speed-up (over \bfastR)},
       xmin=0, xmax=1100,
       ymin=0, ymax=10000000,
       xtick={100,200,300,400,500,600,700,800,900,1000},
       legend pos=north west,
       ymode=log,
       xticklabel style={rotate=45},
       x label style={at={(axis description cs:0.5,-0.05)},anchor=north},
   ]
    
   \addplot[color=orange,mark=diamond,]
       coordinates {(100,11.6884714065)
                    (200,11.4356261736)
                    (300,11.4440296389)
                    (400,11.4917814907)
                    (500,11.5532874866)
                    (600,11.6519783932)
                    (700,11.2425080937)
                    (800,11.248754394)
                    (900,11.3247366131)
                    (1000,11.5092798977)};
    
   \addplot[color=blue,mark=square,]
       coordinates {(100,3402.15251369)
                    (200,3384.73336961)
                    (300,3425.52923667)
                    (400,3407.28618765)
                    (500,3223.09514845)
                    (600,3094.85512799)
                    (700,2963.01456054)
                    (800,2838.22886252)
                    (900,2886.14114256)
                    (1000,2967.15423196)};
    
   \addplot[color=green,mark=triangle,]
       coordinates {(100,20608.4092188)
                    (200,21505.2163824)
                    (300,21915.7190894)
                    (400,22315.6724241)
                    (500,22265.6235532)
                    (600,22491.5420255)
                    (700,21790.2113764)
                    (800,21834.8600227)
                    (900,21956.9077582)
                    (1000,21240.9450613)};
       
   \legend{\bfastPython,\bfastcpu,\bfastgpu}
   \end{axis}
  \end{tikzpicture}
  }
  }
 \end{subfigure}
 \caption{Figure~(a) shows the runtimes for all four implementations considered. Figure~(b) only shows \bfastcpu and \bfastgpu for an easier comparison between these two. Figure~(c) sketches the resulting speed-ups over \bfastR. It can be seen that the many-core implementation is up to four orders of magnitude faster than the existing \ttt{R} implementation and about three orders of magnitude faster than \bfastPython, which depicts a direct implementation of the \bfast~approach. Further, it also yields a speed-up of about eight over the corresponding multi-threaded CPU execution (\bfastcpu), which shows the benefits of using GPUs~for conducting large-scale \bfast~analyses.}
 \label{fig:timings}
\end{figure*}

As shown below, the \bfastgpu~implementation yields significant speed-ups over the existing \ttt{R} implementation \bfastR. This is, among other things, due to the adapted computations (i.e., all time series are treated simultaneously), the fact that fewer (expensive) function calls are needed (in particular, no high-level \ttt{R} function calls), and potential overhead done by \bfastR (\eg, sanity checks). We would like to point out that this implementation was developed with a focus on providing printed output, summaries, and visualizations; it was not optimized for a parallel application for large satellite image time series scenarios, as it is the focus of this work. Nevertheless, it depicts one of the main tools used for this type of analysis in remote sensing and is therefore included in our experimental evaluation. The \bfastPython implementation, however, should give a good intuition for the runtimes needed in case all time series are handled individually.

\subsection{Runtime Analysis}

We start by analyzing the runtimes of all \bfast implementations. For these experiments, we resort to artificial datasets, which are generated as follows:
Each of the $m$ time series is generated by a process using a sinus curve and for half of the time series, a constant will be added to the last 40\% of the corresponding values to make sure that these time series exhibit a break. In addition, some noise is added to all elements of a time series. Generating a single element $y_t$ at time $t$ in a time series is done via 
\begin{equation}
\label{eq:randomgen}
y_t = 0.05 \times \sin\left(\frac{2t\pi}{f}\right) + \epsilon_t + c,
\end{equation}
where $\epsilon_t$ is a small random number and $c$ the constant added to the last 40\% of the time series that should have a break. 
\subsubsection{Influence of m}
We start by investigating the runtime behavior w.r.t. the number $m$ of time series. The other parameters are set to fixed values ($n=100$, $f=23$, $h=50$, $k=3$, $\alpha=0.05$). For this experiment, we vary $m$ from $100,000$ to $1,000,000$ in steps of $100,000$ with time series of length $N=200$. 

The output is shown in Figure~\ref{fig:timings}. It can be seen that the many-core implementation is significantly faster than both \bfastR and \bfastPython. Further, it is also about a magnitude faster than the multi-threaded execution (\bfastcpu, using 4 threads), which depicts a valuable speed-up as well. As expected, the speed-ups for all versions are more or less constant across all sizes of input, see Figure~\ref{fig:timings}~(c).

\subsubsection{Multi- vs. Many-Core}
Next, we focus on a direct runtime comparison between \bfastcpu and \bfastgpu. Five different parts of the \bfastcpu~and \bfastgpu~versions have been timed (the remaining parts are not included because their runtimes are not significant). For \bfastcpu, we focus on the following five phases:
\begin{enumerate}[label=\textbf{\arabic*}]
  \item Create model
  \item Calculate predictions
  \item Calculate residuals
  \item Calculate \mosums
  \item Detect breaks
\end{enumerate}

Accordingly, we have timed five parts of \bfastgpu. Note that these parts are not in a one-to-one correspondence with the ones of \bfastcpu due to the fact that the computation of the residuals and the \mosums have been fused in the \GPU~implementation (see Algorithm~\ref{alg:moving_sums_gpu}). Further, there are additional transfer phases for \bfastgpu:
\begin{enumerate}[label=\textbf{\arabic*}]
  \item Transfer seasonal matrix and dataset to GPU memory
  \item Create model
  \item Calculate predictions
  \item Calculate \mosums
  \item Detect breaks
\end{enumerate}

The runtime experiments are set up in the same way as before. The outcome for the analysis of $m=1,000,000$ time series is shown in Figure~\ref{fig:phases}: It can be seen that, for \bfastcpu, there is not a single computational bottleneck. Instead, the total runtime is spread over all the different phases. This shows that it is actually necessary to address all these parts when designing an efficient massively-parallel version (\eg, both the computation of the moving sums and of the breaks have to be accelerated to achieve a significant speed-up for \bfastgpu). In contrast, there is only one major phase dominating the runtime for \bfastgpu, namely the transfer phase. Regarding the remaining four parts, calculating the \mosums is the biggest one. While it might still be possible to optimize the corresponding kernel (\eg, better memory access that takes advantage of L1/L2 caching), the runtime of this phase is already much smaller than the one for the transfer phase. Hence, aiming at further runtime improvements, the transfer phase has to be addressed next (see below).


\begin{figure}[t]
 \centering
  \begin{subfigure}[\bfastcpu]{
   \centering
   \resizebox{0.35\textwidth}{!}{
   \begin{tikzpicture}%
    \begin{axis}[%
       symbolic x coords={Model,Predictions,Residuals,MOSUM,Breaks},%
       ylabel={runtime (seconds)},%
       ymin=0,%
       xtick=data,%
       xticklabel style={rotate=45},%
     ]%
     \addplot[ybar,bar width=1cm,fill=blue!40!white]%
         coordinates {(Model,1.43755507469)%
                      (Predictions,0.796591997147)%
                      (Residuals,0.710953950882)%
                      (MOSUM,0.503683805466)%
                      (Breaks,0.51752281189)};%
    \end{axis}%
   \end{tikzpicture}%
   }
   }
  \end{subfigure}
  \begin{subfigure}[\bfastgpu]{
   \centering
   \resizebox{0.35\textwidth}{!}{
   \begin{tikzpicture}%
    \begin{axis}[%
       symbolic x coords={Transfer,Model,Predictions,MOSUM,Breaks},%
       ylabel={runtime (seconds)},%
       ymin=0,%
       xtick=data,%
       xticklabel style={rotate=45},%
     ]%
     \addplot[ybar,bar width=1cm,fill=blue!40!white]%
         coordinates {(Transfer,0.445647001266)%
                      (Model,0.00848984718323)%
                      (Predictions,0.0105481147766)%
                      (MOSUM,0.0556058883667)%
                      (Breaks,0.00950598716736)};%
    \end{axis}%
   \end{tikzpicture}%
   }
   }
  \end{subfigure}
 \caption{Runtimes of the five most significant phases of (a) \bfastcpu and (b) \bfastgpu. For the many-core version, the runtimes of all phases have been greatly reduced resulting in the transfer of data between host and device being the remaining computational bottleneck (which cannot be avoided since the image data have to be transferred).}
 \label{fig:phases}
\end{figure}
\begin{figure}[t]
 \centering
 \begin{subfigure}[\bfastcpu]{
  \centering
  \resizebox{0.35\textwidth}{!}{
  \begin{tikzpicture}
   \begin{axis}[
       xlabel={number of time series ($\times 10^3$)},
       ylabel={runtime (seconds)},
       xmin=0, xmax=1100,
       ymin=0,
       xtick={100,200,300,400,500,600,700,800,900,1000},
       legend pos=north west,
       xticklabel style={rotate=45},
       x label style={at={(axis description cs:0.5,-0.05)},anchor=north},
   ]
    
   \addplot[color=red,mark=star,]
       coordinates {(100,0.101739168167)
                    (200,0.205857992172)
                    (300,0.312903165817)
                    (400,0.424304008484)
                    (500,0.538087844849)
                    (600,0.665899991989)
                    (700,0.894320011139)
                    (800,1.11470413208)
                    (900,1.27126908302)
                    (1000,1.43755507469)};
    
   \addplot[color=orange,mark=diamond,]
       coordinates {(100,0.0801391601562)
                    (200,0.160482168198)
                    (300,0.239304065704)
                    (400,0.320423126221)
                    (500,0.398840904236)
                    (600,0.480622053146)
                    (700,0.557192087173)
                    (800,0.638704061508)
                    (900,0.720915079117)
                    (1000,0.796591997147)};
    
   \addplot[color=yellow,mark=*,]
       coordinates {(100,0.0717780590057)
                    (200,0.144429922104)
                    (300,0.212167024612)
                    (400,0.290199995041)
                    (500,0.350390911102)
                    (600,0.436717987061)
                    (700,0.499402999878)
                    (800,0.686568021774)
                    (900,0.678073167801)
                    (1000,0.710953950882)};
    
   \addplot[color=blue,mark=square,]
       coordinates {(100,0.0617299079895)
                    (200,0.10817193985)
                    (300,0.152099132538)
                    (400,0.207213878632)
                    (500,0.258831977844)
                    (600,0.306346178055)
                    (700,0.348140954971)
                    (800,0.407896995544)
                    (900,0.453059911728)
                    (1000,0.503683805466)};
    
   \addplot[color=green,mark=triangle,]
       coordinates {(100,0.0411529541016)
                    (200,0.0800800323486)
                    (300,0.121896028519)
                    (400,0.15827703476)
                    (500,0.20120382309)
                    (600,0.250025987625)
                    (700,0.288776874542)
                    (800,0.349292993546)
                    (900,0.399351119995)
                    (1000,0.51752281189)};
       
   \legend{Model,Predictions,Residuals,MOSUM,Breaks}
   \end{axis}
  \end{tikzpicture}
  }
  }
 \end{subfigure}
 \begin{subfigure}[\bfastgpu]{
  \centering
  \resizebox{0.35\textwidth}{!}{
  \begin{tikzpicture}
   \begin{axis}[
       xlabel={number of time series ($\times 10^3$)},
       ylabel={runtime (seconds)},
       xmin=0, xmax=1100,
       ymin=0,
       xtick={100,200,300,400,500,600,700,800,900,1000},
       legend pos=north west,
       xticklabel style={rotate=45},
       x label style={at={(axis description cs:0.5,-0.05)},anchor=north},
   ]
    
   \addplot[color=red,mark=star,]
       coordinates {(100,0.0471520423889)
                    (200,0.0928900241852)
                    (300,0.136106967926)
                    (400,0.178799152374)
                    (500,0.224735021591)
                    (600,0.266371965408)
                    (700,0.313040018082)
                    (800,0.360351800919)
                    (900,0.40610408783)
                    (1000,0.445647001266)};
    
   \addplot[color=orange,mark=diamond,]
       coordinates {(100,0.00198411941528)
                    (200,0.0026650428772)
                    (300,0.00330114364624)
                    (400,0.00399494171143)
                    (500,0.00413012504578)
                    (600,0.00482106208801)
                    (700,0.00545501708984)
                    (800,0.0062530040741)
                    (900,0.00648403167725)
                    (1000,0.00848984718323)};
    
   \addplot[color=yellow,mark=*,]
       coordinates {(100,0.00149893760681)
                    (200,0.0024950504303)
                    (300,0.00360703468323)
                    (400,0.00461411476135)
                    (500,0.00527405738831)
                    (600,0.00656986236572)
                    (700,0.00719118118286)
                    (800,0.00845384597778)
                    (900,0.00919795036316)
                    (1000,0.0105481147766)};
    
   \addplot[color=blue,mark=square,]
       coordinates {(100,0.00623488426208)
                    (200,0.011234998703)
                    (300,0.016450881958)
                    (400,0.0216388702393)
                    (500,0.0266509056091)
                    (600,0.0318419933319)
                    (700,0.0369877815247)
                    (800,0.0428020954132)
                    (900,0.0477051734924)
                    (1000,0.0556058883667)};
    
   \addplot[color=green,mark=triangle,]
       coordinates {(100,0.00133895874023)
                    (200,0.00230598449707)
                    (300,0.00371599197388)
                    (400,0.00432801246643)
                    (500,0.00456404685974)
                    (600,0.00541806221008)
                    (700,0.00627303123474)
                    (800,0.00770998001099)
                    (900,0.00799107551575)
                    (1000,0.00950598716736)};
       
   \legend{Transfer,Model,Predictions,MOSUM,Breaks}
   \end{axis}
  \end{tikzpicture}
  }
  }
 \end{subfigure}
 \caption{Runtime analysis of the different phases for both the (a) CPU and (b) GPU implementation. It can be seen that all phases of the GPU~implementation have been significantly reduced leading to the transfer of data being the remaining bottleneck.}
 \label{fig:phases-expanded}
\end{figure}
The behavior of the runtimes for the different phases given an increasing amount $m$ of time series is shown in Figure~\ref{fig:phases-expanded}. The previous points derived from Figure~\ref{fig:phases} are still true for these other sizes of datasets: The phases of \bfastcpu~all play a significant part in the total runtime, but for \bfastgpu, it is basically only the phase of transferring the data from host to device that matters. Note that, since the transfer of data cannot be avoided, the many-core implementations for the remaining four parts do not need to be tuned for efficiency anymore (accelerating these phase will not result in any further significant speed-ups).




\subsubsection{Influence of k}

\begin{figure}[t]
 \centering
 \begin{subfigure}[\bfastcpu]{
  \centering
  \resizebox{0.35\textwidth}{!}{
  \begin{tikzpicture}
   \begin{axis}[
       xlabel={$k$},
       ylabel={runtime (seconds)},
       xmin=0, xmax=6,
       ymin=0, ymax=3,
       xtick={1,2,3,4,5},
       legend pos=north west,
       x label style={at={(axis description cs:0.5,-0.05)},anchor=north},
   ]
    
   \addplot[color=red,mark=star,]
       coordinates {(1,1.3674390316)
                    (2,1.40005278587)
                    (3,1.38624691963)
                    (4,1.42948913574)
                    (5,1.43888902664)};
    
   \addplot[color=orange,mark=diamond,]
       coordinates {(1,0.793162107468)
                    (2,0.800498962402)
                    (3,0.806082963943)
                    (4,0.806141138077)
                    (5,0.812373161316)};
    
   \addplot[color=yellow,mark=*,]
       coordinates {(1,0.706356048584)
                    (2,0.707882165909)
                    (3,0.709186077118)
                    (4,0.706754922867)
                    (5,0.706424951553)};
    
   \addplot[color=blue,mark=square,]
       coordinates {(1,0.495589017868)
                    (2,0.504349946976)
                    (3,0.495844841003)
                    (4,0.503814935684)
                    (5,0.499783992767)};
    
   \addplot[color=green,mark=triangle,]
       coordinates {(1,0.504801034927)
                    (2,0.506778001785)
                    (3,0.505088090897)
                    (4,0.504054069519)
                    (5,0.504167079926)};
       
   \legend{Model,Predictions,Residuals,MOSUM,Breaks}
   \end{axis}

  \end{tikzpicture}
  }
  }
 \end{subfigure}
 \begin{subfigure}[\bfastgpu]{
  \centering
  \resizebox{0.35\textwidth}{!}{
  \begin{tikzpicture}
   \begin{axis}[
       xlabel={$k$},
       ylabel={runtime (seconds)},
       xmin=0, xmax=6,
       ymin=0, ymax=1,
       xtick={1,2,3,4,5},
       legend pos=north west,
       x label style={at={(axis description cs:0.5,-0.05)},anchor=north},
   ]
    
   \addplot[color=red,mark=star,]
       coordinates {(1,0.448664903641)
                    (2,0.449517011642)
                    (3,0.447811126709)
                    (4,0.448140144348)
                    (5,0.447113037109)};
    
   \addplot[color=orange,mark=diamond,]
       coordinates {(1,0.00839304924011)
                    (2,0.0070219039917)
                    (3,0.00813603401184)
                    (4,0.00745797157288)
                    (5,0.00694608688354)};
    
   \addplot[color=yellow,mark=*,]
       coordinates {(1,0.00943422317505)
                    (2,0.00958395004272)
                    (3,0.0106711387634)
                    (4,0.0106930732727)
                    (5,0.0112779140472)};
    
   \addplot[color=blue,mark=square,]
       coordinates {(1,0.0546939373016)
                    (2,0.0527069568634)
                    (3,0.0527899265289)
                    (4,0.0526690483093)
                    (5,0.0527081489563)};
    
   \addplot[color=green,mark=triagle,]
       coordinates {(1,0.0095100402832)
                    (2,0.00882792472839)
                    (3,0.00944089889526)
                    (4,0.00882601737976)
                    (5,0.00883388519287)};
       
   \legend{Transfer,Model,Predictions,MOSUM,Breaks}
   \end{axis}

  \end{tikzpicture}
  }
  }
 \end{subfigure}
 \caption{Runtimes for both \bfastcpu and \bfastgpu given different assignment for $\boldsymbol{k}$. It can be seen that $\boldsymbol{k}$ has no significant impact on any of the phases in both versions.}
 \label{fig:k-impact}
\end{figure}
Next the impact of $k$ on the total runtime of \bfastcpu and \bfastgpu is analyzed. We will measure the runtime of the five phases of both versions mentioned above. Only the model creation and prediction phases are expected to be influenced by $k$. Again, we resort to $m=1,000,000$ time series each having length $N=200$ and perform the analysis with different values of $k$ ($k=1,\dots,5$). These are realistic values of $k$ (typically, the value for $k$ is very small, such as $k=3$ or $k=4$). The other settings are as follows: $n=100$, $f=23$, $h=50$, and $\alpha=0.05$.

In Figure~\ref{fig:k-impact}, the impact of $k$ on the runtime of the different phases is shown. One can see that no phases in any version are impacted in a significant way by the value of $k$. This is in line with the runtime/memory requirements of the implementation: For \bfastgpu, the transfer of $\mathcal{O}(Nk + Nm)$ data from host to device is mainly influenced by $m \gg k$ and, since the transfer of data dominates the overall runtime, we do not expect a significant influence of $k$. For \bfastcpu, the parameter $k$ does not influence the computation of the predictions, residuals, \mosums, and breaks from a theoretical perspective as well. It influences the runtime of the model construction, but the practical influence is only small (likely due to the fact of $k$ being too small).


\subsubsection{Influence of h}
\begin{figure}[t]
 \centering
 \begin{subfigure}[\bfastcpu]{
  \centering
  \resizebox{0.35\textwidth}{!}{
  \begin{tikzpicture}
   \begin{axis}[
       xlabel={$h$},
       ylabel={runtime (seconds)},
       xmin=0, xmax=125,
       ymin=0, ymax=5.2,
       xtick={25,50,100},
       legend pos=north west,
       x label style={at={(axis description cs:0.5,-0.05)},anchor=north},
   ]
    
   \addplot[color=red,mark=star,]
       coordinates {(25,3.89206504822)
                    (50,3.91252994537)
                    (100,3.94609808922)};
    
   \addplot[color=orange,mark=diamond,]
       coordinates {(25,0.486392974854)
                    (50,0.494993209839)
                    (100,0.533621072769)};
       
   \legend{Total,MOSUM}
   \end{axis}

  \end{tikzpicture}
  }
  }
 \end{subfigure}
 \begin{subfigure}[\bfastgpu]{
  \centering
  \resizebox{0.35\textwidth}{!}{
  \begin{tikzpicture}
   \begin{axis}[
       xlabel={$h$},
       ylabel={runtime (seconds)},
       xmin=0, xmax=125,
       ymin=0, ymax=0.8,
       xtick={25,50,100},
       legend pos=north west,
       x label style={at={(axis description cs:0.5,-0.05)},anchor=north},
   ]
    
   \addplot[color=red,mark=star,]
       coordinates {(25,0.531475067139)
                    (50,0.52701997757)
                    (100,0.531201124191)};
    
   \addplot[color=orange,mark=diamond,]
       coordinates {(25,0.0526859760284)
                    (50,0.0526609420776)
                    (100,0.0555670261383)};
       
   \legend{Total,MOSUM}
   \end{axis}

  \end{tikzpicture}
  }
  }
 \end{subfigure}
 \caption{Total runtime and runtime of MOSUM phase of (a) \bfastcpu~and (b) \bfastgpu~on different values of $\boldsymbol{h}$. The value of $\boldsymbol{h}$ clearly has no impact on the runtimes.}
 \label{fig:h-impact}
\end{figure}

Finally, we examine the impact of $h$ on the total runtime. Only the \mosum phase and the total runtime will be measured since $h$ can only influence the calculation of the \mosums. Measuring the total runtime serves as a sanity check. Again, we resort to a dataset with $m=1,000,000$ time series of length $N=200$ and make use of different assignments for the parameter $h$ ($25$, $50$, and $100$). The other settings are as follows: $n = 100$, $f=23$, $k=3$, and $\alpha=0.05$.

In general, we would not expect $h$ to have a very large impact on the \mosum phase since only the first sum computed actually uses~$h$. All sums except for the first one are computed from the previous one, so the size of the window does not influence this. In Figure~\ref{fig:h-impact}, we can see the runtime of the \mosum and total runtime and how they are impacted by $h$ for both \bfastcpu~and \bfastgpu. We can clearly see, as expected, that the runtime is not affected by the values of~$h$. We can therefore pick the value of $h$ that best fits our analysis without having to think about the potential impact on the total runtime.

\subsection{Large-Scale Break Detection}

\begin{figure*}[t]%
 \centering%
 \begin{subfigure}[]{%
  \centering%
  \resizebox{0.21\textwidth}{!}{\includegraphics{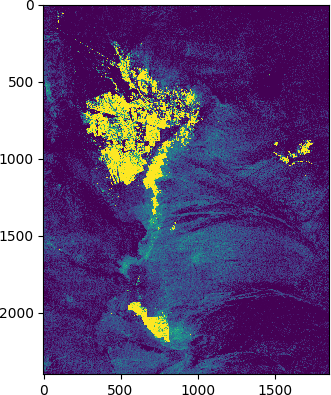}}%
  }%
 \end{subfigure}
 \begin{subfigure}[]{%
  \centering%
  \resizebox{0.21\textwidth}{!}{\includegraphics{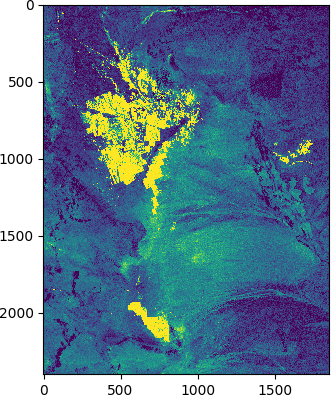}}%
  }%
 \end{subfigure}
 \begin{subfigure}[]{%
  \centering%
  \resizebox{0.21\textwidth}{!}{\includegraphics{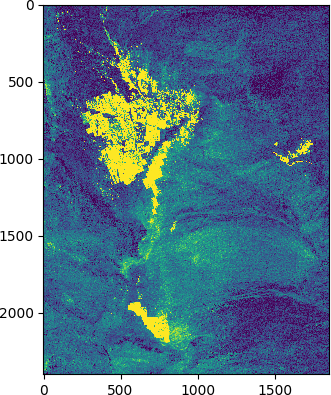}}%
  }%
 \end{subfigure}
 \begin{subfigure}[]{%
  \centering%
  \resizebox{0.266\textwidth}{!}{\includegraphics{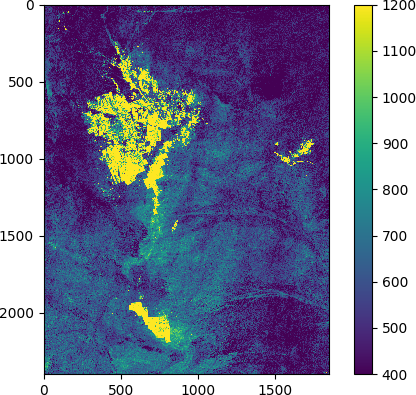}}%
  }%
 \end{subfigure}
 \begin{subfigure}[]{%
  \centering%
  \resizebox{0.21\textwidth}{!}{\includegraphics{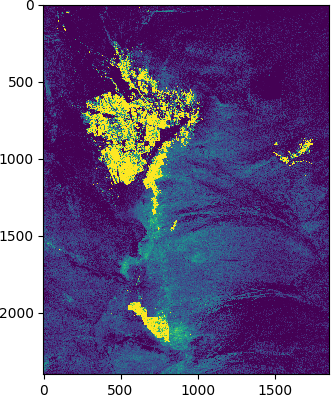}}%
  }%
 \end{subfigure}
 \begin{subfigure}[]{%
  \centering%
  \resizebox{0.21\textwidth}{!}{\includegraphics{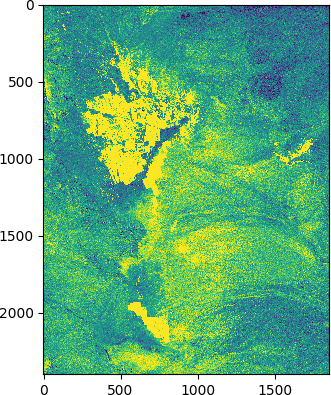}}%
  }%
 \end{subfigure}
 \begin{subfigure}[]{%
  \centering%
  \resizebox{0.21\textwidth}{!}{\includegraphics{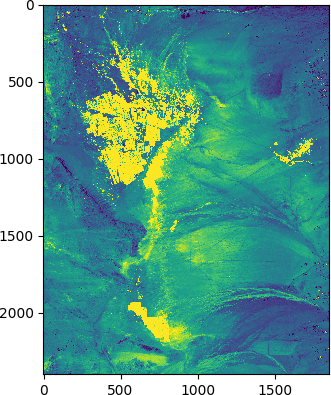}}%
  }%
 \end{subfigure}
 \begin{subfigure}[]{%
  \centering%
  \resizebox{0.266\textwidth}{!}{\includegraphics{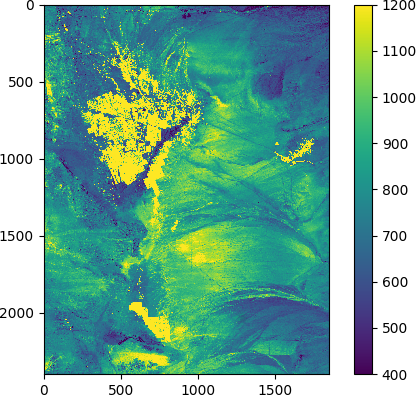}}%
  }%
 \end{subfigure}
 \caption{The (a) 1st, (b) 40th, (c) 80th, (d) 120th, (e) 160th, (f) 200th, (g) 240th, and (h) 288th image of the Chile dataset in a blue/yellow heatmap. Is is evident that between images (e) and (f), many pixels change significantly. This indicates that there are breaks in most of the induced time series.}%
 \label{fig:chile-heatmap-multiple}%
\end{figure*}%

\begin{figure}[t]
  \centering
  \resizebox{0.8\columnwidth}{!}{
  \begin{tikzpicture}
   \begin{axis}[
       xlabel={number of pixels},
       ylabel={runtime (seconds)},
       xmin=0, xmax=7,
       ymin=0, ymax=35,
       xtick={1,2,3,4,5,6},
       xticklabels={$1851\times400\times1$,$1851\times400\times2$,$1851\times400\times3$,$1851\times400\times4$,$1851\times400\times5$,$1851\times400\times6$},
       legend pos=north west,
       xticklabel style={rotate=45},
       x label style={at={(axis description cs:0.5,-0.20)},anchor=north},
   ]
    
   \addplot[color=red,mark=star,]
       coordinates {(1,3.94116711617)
                    (2,9.32816410065)
                    (3,15.6123220921)
                    (4,21.2714140415)
                    (5,27.2735340595)
                    (6,32.7450990677)};
    
   \addplot[color=orange,mark=diamond,]
       coordinates {(1,0.569360971451)
                    (2,1.12266206741)
                    (3,1.72365403175)
                    (4,2.51708102226)
                    (5,3.15297913551)
                    (6,3.88919711113)};
       
   \legend{CPU,GPU}
   \end{axis}

  \end{tikzpicture}
  }
 \caption{Runtimes for parts of the Chile dataset. As expected, one can see that the runtime grows linearly. 
 }
 \label{fig:chile-runtime}
\end{figure}
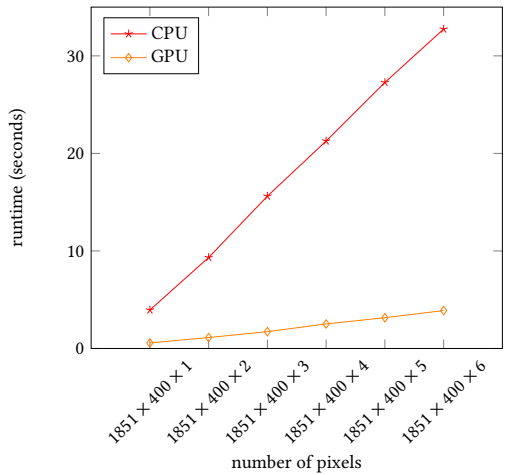

The runtime analysis conducted above clearly shows the benefits of the \bfastgpu~implementation. From a practical perspective, this new implementation makes it possible to process and analyze significantly larger datasets as with the original implementation. In the remainder of this section, we consider a real-world dataset to demonstrate the benefits of our framework.

We tested our \CPU~and \GPU~implementations on a dataset of Landsat imagery over an area of 400,000 ha in the Atacama Desert, Chile~(20.5S, 69.5W), where vegetation dynamics can be observed in a plantation forest. This area was chosen as it is the driest non-polar desert in the world, providing the maximum number of cloud free observations possible over a vegetated area. The dataset was directly acquired from the USGS\footnote{\url{https://earthexplorer.usgs.gov}}, for the P01R74 scene, contains 288 Landsat Collection 1 Tier 1 Surface Reflectance derived NDVI images, starting from 18/01/2000 to 20/08/2017 from different sensors (Landsat 5, 7 Slc-on, and 8). Our dataset is a subset of 2400 $\times$ 1851 pixels in the south-west part of the scene. In Figure~\ref{fig:chile-heatmap-multiple}, a heatmap of a selection of the 288 images is given. It is clear that something happens between the fifth (e) and sixth (f) image, making the break detection analysis relevant for this dataset.

For this analysis, we assign the following values to the involved parameters: $n=144$, $f=365$, $h=72$, $k=3$, $\alpha=0.05$. Note that the time series data are not sampled evenly over the different years. For this reason, one needs to adapt the processing slightly such that one uses the day (number) per year instead of the index $t$ for Equation~(\ref{eq:modelpure}). 

The dataset has been split into six parts of equal sizes, with each analysis performed on one chunk, two chunks and so forth up to the entire data set. The runtimes for the experiment can be seen in Figure~\ref{fig:chile-runtime}, and as expected we see the runtime grows in the same manner as when we performed a similar analysis on artificial datasets. The total runtime for \bfastcpu and \bfastgpu was $32.8$ seconds and $3.9$ seconds, respectively. Note that using the original \texttt{R} code takes about 20 hours. For the sake of comparison, we also considered an additional more powerful multi-core system with 36 logical cores (two Intel(R) Xeon(R) CPUs E5-2666 v3 @ 2.90GHz) and 60GiB of RAM running Ubuntu 16.04.3 LTS. Using this significantly stronger multi-core system, processing this scene took about 5,540 seconds using \bfastR. This is still about 1,000 times slower than \bfastgpu~executed using our main test system.

\begin{figure}[t]
 \centering
  \resizebox{0.75\columnwidth}{!}{\includegraphics{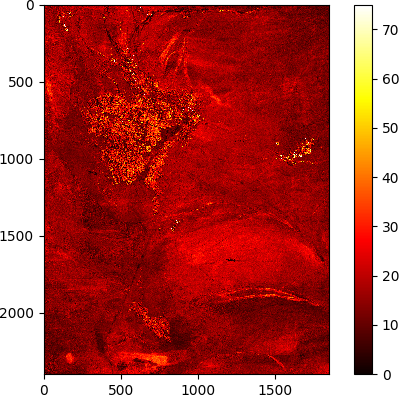}}
 \caption{Heatmap of the maximum absolute values of the pixels' \mosum processes. This map indicates that the breaks had a bigger magnitude in the ``hotter areas''. 
 }
 \label{fig:chile-heatmap}
\end{figure}

As expected, \bfast detected breaks for almost all pixels (more than $99\%$). In Figure~\ref{fig:chile-heatmap}, a heatmap of the maximum absolute values of the pixels' \mosums is shown. It can be seen that some breaks exhibit a larger magnitude for certain regions. Almost all pixels have a break, as the boundary detecting a break is at $2.39$. The spotty areas are the plantation forest, where the breaks are both negative and positive, which is normal as some parts of the forest are being planted (yellow/red) while others are harvested (dark red). The desert areas also experience change, but at a much smaller magnitude than the forest.

\section{Conclusion}
\label{section:conclusions}
In this work, we present a many-core implementation for the \bfast~approach, which depicts one of the state-of-the-art scehmes for change detection in remote sensing. Our implementation is up to four orders of magnitudes faster than the commonly used \ttt{R} implementation, up to three orders of magnitudes faster than a direct \CPU~implementation, and up to ten times faster than a highly-tuned multi-threaded \CPU~execution. Our new implementation can be used to handle significantly larger datasets. In our experimental evaluation, we considered large satellite image time series datasets, which could be processed in a couple of seconds only compared to hours using the original \ttt{R} implementation. 



The current computational bottleneck of the many-core implementation is the transfer of data from host to device. In future, it would be interesting to investigate if the time for transferring the data could be brought down by, \eg, compressing the data prior to transferring it or by investigating the minimal amount of precision needed for an analysis in order to reduce the data volume. It would also be interesting to see if related change detection methods could benefit in a similar fashion from massively-parallel devices as the \bfast approach considered in this work. This also holds true for variants being capable of dealing with many missing values being present in the individual time series. 

\begin{acks}
We would like to thank the anonymous reviewers for their helpful comments and suggestions. Fabian Gieseke acknowledges support from the Danish Industry Foundation through the \emph{Industrial Data Analysis Service} (IDAS).
\end{acks}

\bibliographystyle{ACM-Reference-Format}
\bibliography{biblio}

\end{document}